\documentclass[12pt]{article}
\usepackage{amsmath}
\usepackage{amsfonts}
\usepackage{amssymb}

\numberwithin{equation}{section}

\def\rf#1{(\ref{eq:#1})}
\def\lab#1{\label{eq:#1}}
\def\bj{{\bar J}}
\def\sj{{\jmath}}
\def\bsj{{\bar \jmath}}

\providecommand*{\pder}[3][]{%
\frac{\partial^{#1}#2}{\partial #3^{#1}}}
\providecommand*{\dder}[3][]{%
\frac{d^{#1}#2}{d #3^{#1}}}
\providecommand*{\iu}%
{\ensuremath{\mathrm{i}\,}}

\begin{document}
\vspace*{-1cm}
\noindent
\vskip .3in

\begin{center}
{\large\bf On the symmetric formulation of the Painlev\'e 
IV equation }
\end{center}
\normalsize
\vskip .4in

\begin{center}
H. Aratyn

\par \vskip .1in \noindent
Department of Physics \\
University of Illinois at Chicago\\
845 W. Taylor St.\\
Chicago, Illinois 60607-7059\\
\par \vskip .3in

\end{center}

\begin{center}
J.F. Gomes and A.H. Zimerman

\par \vskip .1in \noindent
Instituto de F\'{\i}sica Te\'{o}rica-UNESP\\
Rua Pamplona 145\\
01405-900 S\~{a}o Paulo, Brazil
\par \vskip .3in

\end{center}

\begin{center}
{\large {\bf ABSTRACT}}\\
\end{center}
\par \vskip .3in \noindent

Symmetries and solutions of the Painlev\'e IV equation 
are presented in an alternative framework which provides 
the bridge between the Hamiltonian formalism 
and the symmetric Painlev\'e IV equation. 
This approach originates from a method developed in 
the setting  of  pseudo-differential Lax formalism  describing
AKNS hierarchy with the Darboux-B\"acklund and Miura transformations.

In the Hamiltonian formalism the Darboux-B\"acklund transformations 
are introduced as maps between solutions of the 
Hamilton equations corresponding to two allowed values of
Hamiltonian's discrete parameter.
The action of the generators of the extended affine Weyl group 
of the $A_2$ root system
is realized in terms of three ``square-roots'' of such 
Darboux-B\"acklund  transformations defined on 
a multiplet of solutions of the Hamilton equations.

\section{Introduction}
\label{section:intro}
The six Painlev\'e equations arise as special scaling limits 
of integrable models and frequently emerge in various 
discrete and continuous models of physics, many-body systems and 
mathematics.  
They were originally proposed by Painlev\'e as the second-order 
differential equations whose solutions have no movable singular 
points except poles. 
The Painlev\'e equations occupy a central place in the study of
nonlinear systems in view of their applications to a variety of 
cross-disciplinary problems. 
Consequently there exists a vast body of literature devoted to a 
remarkable connection between these equations and wide-range of 
problems of mathematical physics.
In the past a lot of effort has been invested in finding all 
rational solutions and their symmetry structure (see f.i. 
\cite{noumi,kajiwara,Clarkson-jmp,Clarkson} for the
case of the Painlev\'e IV equation).

Our work is devoted towards constructing universal 
approach to deal with symmetries and solutions of the Painlev\'e 
equations originating from methods we have previously introduced 
for a class of integrable models described by 
the pseudo-differential Lax operators encountered in the 
Kadomtsev-Petviashvili type of hierarchies. 
Recently, in \cite{AGZ2009}, we reported progress in utilizing
techniques  of integrable models for the purpose of 
obtaining and classifying solutions of the Painlev\'e IV equations. 
In particular, reference \cite{AGZ2009} 
 explored the connection between 
the pseudo-differential Lax hierarchies describing 
various versions of the AKNS hierarchy subject to the string 
equation and the Painlev\'e IV equation for the purpose of
reproducing in a novel way the rational solutions of the 
Painlev\'e IV equation through the Darboux-B\"acklund (DB) transformations 
of the reduced integrable hierarchy.

In this paper we use a generalization of Okamoto's Hamiltonian system
\cite{okamoto}, which explicitly depends on 
four parameters. Three of the parameters, $(v_i,v_j,v_k)$, 
labeled by distinct integers $i,j,k=1,2,3$,  
satisfy the condition $\sum_n v_n=0$ and are associated with the 
root system of the $A_2$ Lie algebra.
A special role is played by, $\epsilon$, the additional fourth 
parameter in the Hamiltonian \rf{hamo}.
For two values of $\epsilon$, $+1$ and $-1$, the Hamilton equations 
reproduce the Painlev\'e IV equation:
\begin{equation} 
y_{xx} = \frac{1}{2y} y_x^2+\frac32 y^3 +4 x y^2 +2(x^2+b)y
-2  \frac{a}{y}
\lab{painyab}
\end{equation}
where $a$ and $b$ are parameters. 
When $\epsilon$ changes its value from e.g $+1$ to $-1$ 
the parameter $b$ shifts by $2$.
In the Hamiltonian setting the Darboux-B\"acklund transformation
is defined as a map between two solutions of the 
Hamilton equations corresponding 
to the two allowed values of the ``epsilon'' parameter.          
In case of the Painlev\'e IV equation  such transformations 
agree with the Darboux-B\"acklund 
transformations of an underlying Lax operator of the reduced  AKNS hierarchy.
In addition to the Darboux-B\"acklund transformations 
we define simple permutation operations which preserve 
the form of the Hamiltonian but permute the $(v_i,v_j,v_k)$
parameters and the corresponding solutions of 
the Hamilton equations. 
Defining Hamiltonian for the Painlev\'e IV equation in the
form which is left invariant under the  permutation operators, allows us 
to study their group symmetries in a natural manner.
The study leads to realization of the Darboux-B\"acklund transformations 
$G_n, n=1,2,3$ as shift operators acting on the parameters according to :
\begin{equation}
G_n (v_l)=v_l+\tfrac13-\delta_{n,l}, \quad n,l=1,2,3\, .
\lab{Gndiag}
\end{equation}
The action of the extended affine Weyl group generators associated with
the roots is realized in terms of the ``square-roots'' $g_n,\;n=1,2,3$ of the 
Darboux-B\"acklund transformations such that $G_n=g_n^2$ for $g_n$ 
defined on a multiplet of solutions of the symmetric Hamilton equations 
and satisfying algebraic relations :
\begin{equation}
g_n g_m g_n= g_m g_n g_m ,\quad (g_n g_m g_n)^2=1,\;\;\;
(g_n^2 g_m)^2=1,\;\;\; ( g_m g_n^2)^2=1,
\lab{grelations}
\end{equation}
for distinct $n,m=1,2,3$.

The results of this study provide insight into the structure of solutions
and connections between different symmetric realizations of the 
Painlev\'e equations. 
There are several advantages of the approach we are proposing. 
First,  it reproduces the symmetric form of the Painlev\'e IV system 
invariant under extended affine Weyl group from a single 
scalar equation for the Jimbo-Miwa-Okamoto $\rho$ function
\cite{jimbo,okamoto} and its simple symmetry structure.
Secondly, the underlying Darboux dressing chain 
described in relations \rf{rhoxyi} and \rf{dchain}
allows for establishing simple explicit 
relations between elements of the symmetric Painlev\'e equation 
and the underlying Jimbo-Miwa-Okamoto $\rho$-function.
This equivalence reduces the task of finding rational solutions of 
the Painlev\'e IV equation to finding polynomial solutions of 
the $\rho$-function equation and acting on them with 
B\"acklund transformations realized as compositions
of permutation operations and Darboux-B\"acklund transformations.

The paper is organized as follows.
In Section \ref{section:akns}, we briefly review the constrained AKNS
hierarchy emphasizing its symmetry under Darboux-B\"acklund 
transformations operating on various pseudo-differential
Lax representations of the hierarchy. We also derive an equation for 
the Jimbo-Miwa-Okamoto function $\rho=(\ln \tau)_x$, for the tau-function
$\tau$, which is fully equivalent to the Painlev\'e IV equation.
In Section \ref{section:symmetryP}, the Hamiltonian approach
is shown to be equivalent with the symmetric Painlev\'e IV formalism.
The ``square-roots'' of the Darboux-B\"acklund transformations are shown to give rise to 
realization of the affine Weyl group symmetry of $A_2$.
 It is also shown how the equation for 
$\rho$ follows from the symmetric Painlev\'e IV equations.
Although material in Section \ref{section:symmetryP} is inspired 
by Section \ref{section:akns} both sections can be read independently.
Finally, in Section \ref{section:applications} the symmetry
of the $\rho$ equation is applied towards construction of rational
solutions of the Painlev\'e IV equation.

\section{AKNS hierarchy, string condition and B\"acklund transformations}
\label{section:akns}
We show in this section how the the pseudo-differential Lax 
formalism of the AKNS hierarchy with the Darboux-B\"acklund 
symmetry augmented by the Virasoro constraint can be used 
to uncover the symmetry 
structure of the Painlev\'e IV equation.
Deriving the Painlev\'e equations as limits of integrable soliton
equations has an advantage that the symmetry transformations
which are present in the original integrable model and which survive
the imposed limit can effectively be used to produce 
all rational solutions of the Painlev\'e equations out of few basic 
polynomial solutions.
Recent insight into close links of the Painlev\'e
equations with the tau function approach \cite{PVI} and even more 
fundamentally with the Lax formalism has opened new ways to tackle the problem of 
uncovering a rich symmetry structure among solutions but also provided 
a direct method to generate solutions of the Painlev\'e equations. 
In particular, the Darboux-B\"acklund transformations of 
integrable models provide a natural framework for the derivation 
of symmetry of the Painlev\'e IV equation \cite{AGZ2009}.

This section is based on two observations put forward in  \cite{AGZ2009}.
First, the Painlev\'e IV equation is obtained as 
reduction of the AKNS hierarchy subject to the additional non-isospectral 
Virasoro symmetry constraint (also known as the string equation).
Next, we recall from \cite{Aratyn:1996nb} that the  
Darboux-B\"acklund transformations commute 
with the additional-symmetry Virasoro flows  
and thus for that reason they induce symmetry operations on
the Painlev\'e IV equation. The Miura map between 
two different Lax realizations of the AKNS hierarchy \cite{Aratyn:1993zi}
is then used to reveal a more detailed structure of solutions and their 
symmetries.
This structure will give rise to an explicit construction of the 
symmetric Painlev\'e IV equation.

\subsection{Reduced AKNS Lax formalism}
\subsubsection{String equation}
To introduce the AKNS Lax formalism with the additional 
Virasoro symmetry flows we define the pseudo-differential Lax 
operator :
\begin{equation}
L =  \partial_x-r\partial_x^{-1}q \, .
\lab{f-5}
\end{equation}
The associated isospectral $t_n$-flow defined through the Lax equation: 
$\partial L/ \partial t_n = \lbrack L^{n }_{+},
L \rbrack $ amounts for $n=2$ to :
\begin{equation}
\pder{}{t_2} q +q_{xx}-2 q^2 r = 0, \quad
\pder{}{t_2} r -r_{xx}+2 q r^2 = 0 ,\quad 
\pder{}{t_2} \rho = 2 \left( -q r_x+r q_x\right)\, .
\lab{akns}
\end{equation}
The first two equations reproduce the conventional  
AKNS equations for the potentials $r$ and $q$. The third 
equation describes a flow of the so-called squared eigenfunction
$\rho$ (see f.i. \cite{squared}), such that $\rho_x   = - 2 r q$.

The AKNS hierarchy  can be augmented  by infinitely many Virasoro symmetry flows
\cite{add-symm}, which commute with the isospectral flows. 
A closed subset of three additional Virasoro flows, which 
forms the $sl(2) $ subalgebra of the Virasoro algebra, has been 
shown to preserve the form of  the Lax operator of the 
AKNS hierarchy \cite{Aratyn:1996nb}. 
As in \cite{AGZ2009}, we reduce the original AKNS
hierarchy by setting one of these flows to zero. That 
amounts to imposing the following ``string equation'':
\begin{equation}
- x q_x -2 t_2 \pder{}{t_2} q = q + \nu q, \, \,  \,
 x r_x +2 t_2 \pder{}{t_2} r = -r + \nu r , \,  \, \,
 x \rho_x +2  t_2 \pder{}{t_2} \rho = - \rho \, ,
\lab{string_eq3}
\end{equation}
where $\nu$ is a free parameter.
One further reduces the hierarchy by 
setting $t_3$ and all the higher flows to zero
and using the AKNS equation \rf{akns} to eliminate the 
$t_2$-dependence from $r, q, \rho$ for the fixed $t_2=-1/4$ value.
This procedure turns equation \rf{string_eq3} into the constraint 
given by
\begin{equation}
\begin{split}
- x q_x +\frac12 \left(-q_{xx}+2q^2r\right) &= q + \nu q ,\quad
 x r_x - \frac12 \left(r_{xx}-2q r^2\right) = -r + \nu r \\
 \rho+ x \rho_x &= \rho-2 x rq = q_x r-q r_x
\end{split}
\lab{string_eq4}
\end{equation}
Few simple algebraic steps followed by an integration
(see \cite{AGZ2009}) yield from eq. \rf{string_eq4} :
\begin{equation}
q_x r_x = q^2 r^2 +2 x \rho -2 \nu rq -(\mu^2-\nu^2)\, ,
\lab{fixc}
\end{equation}
with $\mu^2-\nu^2$ emerging as an integration constant with 
a new parameter $\mu$. In what follows $\mu$
together with $\nu$ will parametrize solutions of 
the Painlev\'e IV equation.

Dividing the first of eqs. \rf{string_eq4} by $q$ and second
by $r$ and summing them and multiplying the result
by $rq$ produces an equation :
\begin{equation}
-x^2 \rho_x +x \rho +\frac14 \rho_{xxx} 
+2 \nu \rho_x + \frac34 \rho^2_x =\mu^2-\nu^2 \, ,
\lab{rho-eq}
\end{equation}
entirely expressed in terms of only one variable $\rho$.
Equation \rf{rho-eq} was obtained after 
inserting $q_x r_x$ from eq. \rf{fixc} and
eliminating the product of $r$ and $q$ through $rq=-\rho_x/2$.

Equation \rf{rho-eq} is a special case of the Chazy I equation 
\cite{cosgrove}. It can be integrated into \cite{jimbo} :
\begin{equation}
 \rho_{xx}^2 = 4 \left(x \rho_x-\rho\right)^2 -2 \rho_x^3
 -8 \nu \rho_x^2 +8 (\mu^2-\nu^2) \rho_x-8 C\, ,
\lab{jmo}
\end{equation}
with $C$ being another integration constant. 
Setting the integration constant $C$ to zero simplifies eq. \rf{jmo}
to : 
\begin{equation}
\bigl( 2 \left(x \rho_x-\rho\right)+\rho_{xx}\bigr)\bigl( 
2 \left(x \rho_x-\rho\right)-
\rho_{xx}\bigr) = 2 \rho_x  \big\lbrack\rho_x-2
 (\mu-\nu)\big\rbrack
\big\lbrack\rho_x+2 (\mu+\nu)\big\rbrack \, .
\lab{jmo2}
\end{equation}
It is well-known (see f.i. \cite{Clarkson-jmp,AGZ2009}) that two 
functions defined as 
\begin{equation} 
\begin{split}
y_{+}&=- \frac{q_x}{q}- 2 x
=  \frac{-1}{2\rho_x}\left(2 \left(x \rho_x-\rho\right)+\rho_{xx}\right)\\
y_{-}&= \frac{r_x}{r}- 2 x=
\frac{-1}{2\rho_x}\left(2 \left(x \rho_x-\rho\right)-\rho_{xx}\right)
\end{split}
\lab{yrho}
\end{equation} 
satisfy the Painlev\'e IV equation \rf{painyab}
with  $b=\nu \pm 1$ and $a= \mu^2$ for $y=y_{\pm}$.

\subsection{The Darboux-B\"acklund transformations}
\label{subsection:DB}
In the context of the  AKNS Lax hierarchy we consider
the Darboux-B\"acklund transformation realized  
as a similarity transformation by an operator 
$T=r \partial_x r^{-1}$ :
\[
L = \partial_x - r \partial_x^{-1} q \;\;\to \;\; {\bar L} = T\,L\,T^{-1}=
\partial_x - {\bar r} \partial_x^{-1} {\bar q}\, .
\]
This transformation is known to leave the AKNS Lax equations invariant.
It is also known that the  DB transformations commute 
with the additional-symmetry Virasoro 
flows  \cite{Aratyn:1996nb} and consequently these transformations
will also leave equation \rf{jmo2} invariant since it was obtained as a 
reduction of the AKNS equation under an additional Virasoro constraint.
Here, we will show  how to use the  DB transformations to transform
solutions of equation \rf{jmo2} to other solutions of these equations
characterized by new values of the parameter $\nu$.

A simple calculation yields 
\begin{equation}
{\bar r} = r (\ln r)_{xx} - r^2 q , \quad {\bar q}=-\frac{1}{r}\, .
\lab{rq-db}
\end{equation}
It is equally easy to formulate the adjoint  Darboux-B\"acklund 
transformation,  
generated by acting with
$S^{*}=q^{-1} \partial_x q$ on the pseudo-differential Lax operator
through the following similarity transformation :
\[
L = \partial_x - r \partial_x^{-1} q \;\,\to \;\, {\widetilde L} = S^{*\,-1}\,L\,
S^{*}= (q^{-1} \partial_x^{-1} q) L (q^{-1} \partial_x q)=
\partial_x - {\widetilde  r} \partial_x^{-1} {\widetilde q}
\]
with
\begin{equation}
{\widetilde q} = -q (\ln q)_{xx} + q^2 r , \quad {\widetilde r}=\frac{1}{q}\, .
\lab{rq-adb}
\end{equation}
It is convenient to rewrite actions of ${^{\sim}}$ and 
$\bar{\phantom{ a}}$ in eqs. \rf{rq-adb} and \rf{rq-db}
as, respectively, transformations $G$ and $G^{-1}$
acting on variables $J,\bj$ defined as :
\[
\bj= -rq =  \rho_x/2, \quad J = (\ln q)_x=-y_{+}-2x\, .
\]
In terms of the above variables the DB transformations
from eqs. \rf{rq-adb} and \rf{rq-db}
take the following form \cite{Aratyn:1993zi} :
\begin{xalignat}{2}
G (J)  & \equiv  J + \left( \ln \left( \bj + J_x \right) \right)_x &
G ( \bj)& \equiv \bj + J_x
\lab{jtransf} \\
G^{-1} (J) &\equiv J - \left( \ln \bj \right)_x  &
 G^{-1} (\bj) &\equiv
\bj + \left( \ln \bj  \right)_{xx} - J_x\, .
\lab{jtransf2}
\end{xalignat}
It follows from \rf{jtransf}-\rf{jtransf2} and relation \rf{yrho} that:
\begin{equation}
G(y) = y- \left(\ln \left( y_x+y^2+2xy+2 \nu +4\right)\right)_x, \;
G^{-1}(y) = y+ \left(\ln \left( y_x-y^2-2xy-2 \nu\right)\right)_x\, ,
\lab{Grho}
\end{equation}
where for notational simplicity we set $y$  for $y_{+}$ 
from relation \rf{yrho}.
The DB transformations $G^{\pm 1}$ agree with Murata's transformations
$T_{\mp}$ \cite{murata} after identifying
parameters $\theta, \alpha$ from \cite{murata} with $\mu, -\nu-1$.
The action of $G^{\pm 1}$  maps the parameters $\mu,\nu$ to
$\mu,\nu \pm 2$.

There also exists a set of variables $\sj,\bsj$ entering 
a different alternative pseudo-differential Lax realization of the AKNS hierarchy.
These variables are related to $J,\bj$ via a 
Miura transformation \cite{Aratyn:1993zi} :
\begin{equation}
J = - \sj \,- \bsj \, + \frac{\sj_x}{\sj}  ; \qquad
\bj = \bsj \, \sj    \, .        \lab{miura1}
\end{equation}
In terms of variables  $\sj,\bsj$ one can define a 
``square-root'' of $G$ transformation as :
\begin{xalignat}{2}
g (\sj\,)  & \equiv  \bsj\, - \frac{\sj_x}{\sj} &
g\;( \bsj\,) &\equiv \sj
\lab{sjtransf} \\
g^{-1} (\bsj\,) &\equiv \sj\, + \frac{ \bsj_x}{ \bsj  }
&g^{-1} ( \sj\,)& \equiv \bsj \, ,
\lab{sjtransf2}
\end{xalignat}
such that the following relation \cite{Aratyn:1993zi} :
\begin{equation} g^2=G\, ,
\lab{gsquare}
\end{equation}
holds when both sides are applied on $J, \bj$ defined by equation \rf{miura1}.

{}From the above relations one obtains simple transformation rules:
\begin{equation}
g(y)=y-\left(\ln (-\sj+y+2x)\right)_x, \;\;\;\,
g^{-1}(y)=y+\left(\ln (\sj)\right)_x\, ,
\lab{ggy}
\end{equation}
which agree in general with transformations derived using the Schlesinger
equations in \cite{fokas}. In the above equation again for notational 
simplicity $y$  denotes $y_{+}$ from relation \rf{yrho}.

The B\"acklund transformation $g$ 
when applied on solutions expressed by $\sj,\bsj$ shifts 
both $\nu$ and $\mu$ by $1$ (see subsection \ref{subsection:hamiltonP}
for details in terms of $v_n$ parameters) in such a way that acting
twice with $g$ agrees with the formula \rf{gsquare}.
Because of the property of  $g$ transformation to shift both parameters 
of the Painlev\'e IV equation this transformation is very useful
in deriving solutions corresponding to new values of 
the parameters \cite{AGZ2009}.

It turns out that both $-\sj$ and $-\bsj$, in addition to $y_{+}$, are
solutions to the Painlev\'e IV equation. In the next section we will present
a systematic approach to dealing with a presence of
all the solutions $y_{\pm}, -\sj$ and $-\bsj$ 
which emerged here from the AKNS hierarchy structure behind
the Painlev\'e IV equation.

\section{Symmetry of the Painlev\'e IV equation}
\label{section:symmetryP}
\subsection{Hamiltonian approach to the Painlev\'e IV equation}
\label{subsection:hamiltonP}
The formalism is defined in terms of the following generalization of
the Painlev\'e IV Hamiltonian from \cite{okamoto} :
\begin{equation}
 H =  2 P^2 Q -\epsilon \left( Q^2 +2 x Q +2 (v_{j}-v_{i}) \right) P +
(v_{k}-v_{i})   Q - 2 v_{i}x\, ,
\lab{hamo}
\end{equation}
where $\epsilon$ is a constant and the parameters $v_{n} , n=1,2,3$
satisfy condition $\sum_{n} v_{n}=0$ and where $i,j,k$ are fixed distinct
numbers between $1$ and $3$.
The resulting Hamilton equations are:
\begin{align}
Q_x &= \pder{H}{P} = 4 Q P -\epsilon \left(Q^2+2 x Q +2  (v_{j}-v_{i}) \right)
     \lab{qx}\\
P_x &=- \pder{H}{Q} = -2 P^2 +\epsilon \left( 2 QP +2 x P \right) - 
(v_{k}-v_{i}) \, .
      \lab{px}    
\end{align}
For two values of $\epsilon$ such that $\epsilon^2=1$ the equation for $Q_{xx}$
derived from \rf{qx}-\rf{px} (after elimination of $P$) can be cast
in the form of the Painlev\'e IV equation \rf{painyab}
with
\begin{equation}
a = (v_{j}-v_{i})^2, \quad  b=- \epsilon -3 v_{k}\, .  
\lab{mub}
\end{equation}  
Due to the Hamilton equations it holds that
\begin{equation}
\dder{}{x} H = H_x 
  =-2 \epsilon Q P -2 v_{i} \, . 
\lab{hx}
\end{equation}
By taking a derivative of \rf{hx}, one obtains 
\[
H_{xx}
  =\epsilon Q \left(H_{x} +2 v_{k}\right) + 2 P \left( H_{x} + 2 v_{j}\right)
  \, . 
\]
Furthermore by combining relation \rf{hx} with the definition of $H$ one finds
that
\begin{equation}
2 \left( x H_{x} -H\right) 
   = - Q \left(H_{x} +2 v_{k}\right) + 2 \epsilon P \left( H_{x} + 2
   v_{j}\right) \, .
\lab{xHH}
\end{equation}
One now easily derives expressions for $Q$ and $P$ in terms of the
Hamiltonian $H$ and it's derivatives :
\begin{equation}
Q^{(k)} =\frac{2 \left( x H_{x} -H\right) -\epsilon H_{xx}}{(-2)\left(H_{x} +2 v_{k}\right) }, \quad
2 P^{(j)}=\frac{2 \left( x H_{x} -H\right) +\epsilon H_{xx}}{(2
\epsilon)\left(H_{x} +2 v_{j}\right) }\, ,
\lab{Q2Pi}
\end{equation}  
where we labeled $Q$ and $P$ by an index  of the $v$ parameter appearing in
their denominators.

Plugging the values of $Q$ and $P$ from expressions
\rf{Q2Pi} into the relation $2 Q P=2 Q^{(k)} P^{(j)}= -\epsilon
\left(H_{x} +2 v_{i}\right)$  
yields
\begin{equation}
\left(2 \left( x H_{x} -H\right) - H_{xx}\right)
\left(2 \left( x H_{x} -H\right) + H_{xx}\right)
 =4 \prod_{n=1}^3
 \left(H_x +2 v_n\right) \, ,
\lab{Hvv}
\end{equation}
Let us define an operation $\pi_{i,k}$  exchanging parameters
$v_i$ and $v_k$ into each other according to :
\begin{equation}
\pi_{i,k} (v_i,v_j,v_k)= (v_k,v_j,v_i) 
\lab{piik}
\end{equation}
for distinct $i,j,k$. Equation \rf{Hvv} is manifestly invariant under 
all three permutations $\pi_{i,j}$ for $\pi_{i,j} (H)=H$.

The permutation $\pi_{i,k}$ transforms $Q^{(k)}$ in expression \rf{Q2Pi} 
to $Q^{(i)}$ by replacing $v_{k}$ with $v_{i}$. 
Accordingly,  $\pi_{i,k}$  changes of the parameters $a,b$ \rf{mub} in 
the  Painlev\'e IV equation \rf{painyab} from 
$a_k=(v_{j}-v_{i})^{2},b_k=-\epsilon
-3 v_{k}$ to $a_i=(v_{k}-v_{j})^{2},b_i=-\epsilon-3 v_{i}$ or
\[
b_{i} = -\frac32  \epsilon -\frac12 b_{k}+\frac32 \eta \sqrt{a_k},
\quad
a_{i} = \left( b_{k} + \epsilon +  \eta \sqrt{a_k} \right)^{2}/4, \;\; \eta = \pm 1
\]
in the form in which these types of the B\"acklund transformations first appeared in 
\cite{lukashevich,kitaev}.

Define function $\rho^{(k)}$ such that
\begin{equation}
 \rho^{(k)}_{x}=2H_{x} +4 v_{k} \, .
\lab{rhodef}
 \end{equation}
Then plugging $\rho=\rho^{(k)}$ into equation \rf{Hvv} 
reproduces eq. \rf{jmo2}, with $\mu$ and $\nu$ given by
$\mu^{2} 
=(v_{j}-v_{i})^2$ and $\nu
=-3 v_{k}$.
Therefore by comparing with eq. \rf{mub} and expression for $Q$ in relation
\rf{Q2Pi} we find that 
\begin{equation}
y_{\pm}= Q^{(k)}_{-\epsilon}= \frac{-1}{2\rho_x}\left(2 \left(x \rho_x-\rho\right)\pm\rho_{xx}\right)
  = \frac{-1}{2\rho_x}\left(2 \left(x \rho_x-\rho\right)- \epsilon
  \rho_{xx}\right),\;\; \epsilon = \mp 1
\lab{yrhok}
\end{equation}
will solve the Painlev\'e IV equation \rf{painyab}
with $a=\mu^{2}$, $b=\nu \pm 1$.

Next, we will study symmetry of equation \rf{jmo2} for the $\rho$-function.
One notes that the left hand side of equation \rf{jmo2}
remains invariant under substitution $\rho={\widetilde \rho} + 
C x$ for any constant $C$.
However for two values of $C$, namely $C=2 (\mu-\nu)$ and 
$C=-2 (\mu+\nu)$,
the right hand side can be given the form $2 {\widetilde \rho}_x  \lbrack {\widetilde \rho}_x-2
({\widetilde \mu}-{\widetilde \nu})\rbrack \lbrack{\widetilde \rho}_x+
2 ({\widetilde \mu}+ {\widetilde \nu})\rbrack$ with
\begin{xalignat}{3}    
{\widetilde \rho}=\rho^{(i)}&= \rho-2 (\mu-\nu)x, & \nu^{(i)}
&= \frac32 \mu- \frac12 \nu, & \mu^{(i)}& = \pm \frac12 (\mu+\nu) \lab{kone}\\
{\widetilde \rho}=\rho^{(j)}&= \rho+2 (\mu+\nu)x, & \nu^{(j)}
&= -   \frac32 \mu-  \frac12  \nu, & \mu^{(j)}& = \pm \frac12 
(\mu-\nu) \, . \lab{ktwo}
\end{xalignat}
Thus for the above two values of constant $C$ the transformation $\rho 
\to  \rho^{(n)}$
takes the ``old'' solution of equation \rf{jmo2} to the ``new'' solution of 
equation \rf{jmo2} with the new parameters  $\mu^{(n)}$,
$\nu^{(n)}$ for $n=i,j$. Note, that  \rf{ktwo} is obtained
from \rf{kone} by $ \mu \to -\mu$. Since $\rho$ is
a solution of equation \rf{jmo2}, which only depends on a parameter $\mu^{2}$
it is therefore invariant under $ \mu \to -\mu$.
One sees by inspection that eq. \rf{jmo2} can be rewritten as
\begin{equation}
\begin{split}
&  \bigl( 2 \left(x \rho^{(n)}_{x}-\rho^{(n)}\right)+\rho^{(n)}_{xx}\bigr)\bigl( 
2 \left(x \rho^{(n)}_{x}-\rho^{(n)}\right)-
\rho^{(n)}_{xx}\bigr) =2 \rho^{(i)}\rho^{(j)}\rho^{(k)}
   \\
&=
2 \rho^{(n)}_{x}  \big\lbrack\rho^{(n)}_{x}-2
 (\mu^{(n)}-\nu^{(n)})\big\rbrack
\big\lbrack\rho^{(n)}_{x}+2 (\mu^{(n)}+\nu^{(n)})\big\rbrack ,\;\;
n=i,j,k \,,
\lab{rho-eqi}
\end{split}
\end{equation}
where we used notation $\rho^{(k)}=\rho, \nu^{(k)}=\nu, \mu^{(k)}=\mu$.
Also, note that $\sum_{i=1}^3 \nu^{(i)}   =0$.

One verifies that $\rho^{(i)}_{x}=2 H_{x}+4v_{i}$ and
$\rho^{(j)}_{ x}=2 H_{x}+4v_{j}$. Recall that $\rho_{x}
=2 H_{x}+4v_{k}=\rho^{(k)}_{x}$. 
On basis of the definition \rf{piik} it follows that
\begin{equation} 
\pi_{k,j} \left(  \mu,\nu \right)= \left(  \mu^{(j)},\nu^{(j)} \right),\quad
\pi_{k,i} \left(  \mu,\nu \right)= \left(  \mu^{(i)},\nu^{(i)} \right)
\lab{pimunu}
\end{equation} 
It also follows that $\pi_{j,i} 
\left(  \mu^{(j)},\nu^{(j)} \right)= \left(  \mu^{(i)},\nu^{(i)} \right)$
and in agreement with the definition \rf{piik} that :
\begin{equation}
\pi_{k,i} (\rho^{(k)})  = \rho^{(i)}, \quad
\pi_{k,j}(\rho^{(k)}) =\rho^{(j)} \, .
\lab{pirho}
\end{equation}

In accordance with \rf{kone} and \rf{ktwo} we now extend definition
\rf{yrhok} of solutions to the Painlev\'e IV equation to :
\begin{equation}
y^{(n)}_{\pm}=  \frac{-1}{2\rho^{(n)}_x}\left(2 \left(x \rho^{(n)}_x-
  \rho^{(n)}\right)\pm\rho^{(n)}_{xx}\right)
,\;\;n=1,2 ,3 \, ,
\lab{yrhoi}
\end{equation}
with $y^{(k)}_{\pm}= y_{\pm}$. 
Substituting in the above definition the expressions of $\rho^{(n)}$ yields
\begin{eqnarray}
y^{(i)}_{\pm}&= Q^{(i)}_{{-\epsilon}}= \dfrac{-1}{2\left(\rho_x-2(\mu-\nu)\right)}
\left(2 \left(x \rho_x-\rho\right)\pm\rho_{xx}\right)
\lab{yrho1}\\
y^{(j)}_{\pm}&=  Q^{(j)}_{{-\epsilon}}= \dfrac{-1}{2\left(\rho_x+2(\mu+\nu)\right)}
\left(2 \left(x \rho_x-\rho\right)\pm\rho_{xx}\right)
\lab{yrho2}
\end{eqnarray}
Note that $y^{(i)}_{+}$ solves the Painlev\'e IV equation
with $\mu^{(i)}, \nu^{(i)}$ from \rf{kone} and
$y^{(i)}_{-}$ solves the Painlev\'e IV equation
with $\mu^{(i)}, \nu^{(i)}-2$.
Similarly, $y^{(j)}_{+}$ solves the Painlev\'e IV equation
with $\mu^{(j)}, \nu^{(j)}$ from \rf{ktwo} and
$y^{(j)}_{-}$ solves the Painlev\'e IV equation
with $\mu^{(j)}, \nu^{(j)}-2$.
Comparing with equation \rf{Q2Pi} one sees that $-2\epsilon P^{(i)}=
y^{(i)}_{\mp},
-2\epsilon P^{(j)}=y^{(j)}_{\mp}$ for the two allowed values of $\epsilon$.
They reproduce functions $-\sj$ and $-\bsj$ found in subsection
\ref{subsection:DB} within the AKNS hierarchy.

Based on relation \rf{pirho} the symmetry operations defined 
in \rf{piik} transform different
solutions $y^{(i)}_{\pm}$ into each other according to 
\begin{equation}
\pi_{i,k} (y^{(i)}_{\pm})= y^{(k)}_{\pm} \, .
\lab{piiky}
\end{equation}

\subsection{Symmetric Painlev\'e IV equations}
\label{subsection:symmeqs}
It is a simple consequence of equation \rf{jmo2} that
\begin{equation}
 y^{(i)}_{\pm} y^{(j)}_{\mp}= \frac12 \rho^{(k)}_{x} , \qquad i,j,k= 1,2,3\, ,
\lab{bil}
\end{equation}
for distinct $i,j,k$.
This equation is manifestly invariant under all three
$\pi_{i,j}$ transformations.

Inserting expressions \rf{hx} and \rf{rhodef} into the Hamiltonian equation 
\rf{qx} yields
\[ \rho_x=-\epsilon Q_x-Q^2-2xQ -2 (-3 v_k)\, ,
\]
which after recalling that $\epsilon=\pm 1$ corresponds to $y^{(n)}_{\mp}$
can be rewritten as :
\begin{equation}
\rho^{(n)}_x = y^{(n)}_{+\,x} -(y^{(n)}_{+})^2 -2 x y^{(n)}_{+} -2 \nu^{(n)}
= -y^{(n)}_{-\,x}-(y^{(n)}_{-})^2-2 x y^{(n)}_{-}-2 \nu^{(n)}, 
\lab{rhoxyi}
\end{equation}
for $n=1,2,3$.

Due to an elementary relation 
\begin{equation}
\rho^{(j)}_x +2 \nu^{(j)}= \frac12 \rho^{(i)}_x+\frac12 \rho^{(k)}_x\, ,
\lab{rho2nu}
\end{equation}
valid for distinct $i,j,k$, 
we get from equations \rf{bil} and \rf{rhoxyi} the following two identities:
\begin{eqnarray}
y^{(j)}_{+} y^{(k)}_{-}+y^{(j)}_{+} y^{(i)}_{-}
&=y^{(j)}_{+\,x} -(y^{(j)}_{+})^2 -2 x y^{(j)}_{+} \lab{darbchain1}\\
y^{(j)}_{-} y^{(k)}_{+}+y^{(j)}_{-} y^{(i)}_{+}
&=-y^{(j)}_{-\,x} -(y^{(j)}_{-})^2 -2 x y^{(j)}_{-} \lab{darbchain2}
\end{eqnarray}
We can summarize the above equations as :
\begin{equation}
y^{(i,\, j)}_{\mp}+ y^{(j)}_{\mp}+y^{(k)}_{\pm}=
-2 x \, , 
\lab{sumsumyik}
\end{equation}
for distinct $i,j,k$ and with
\[
y^{(i,\,j)}_{\mp} =   y^{(i)}_{\pm}\pm \left( \ln
\left(y^{(j)}_{\mp}\right)\right)_x \, .
\]
Consider one of equations in \rf{sumsumyik}, e.g. $y^{(i,\, j)}_{-}+ y^{(j)}_{-}+y^{(k)}_{+}=
-2 x $ together with 
\[
y^{(i)}_{+} y^{(j)}_{-}=  \left(  y^{(i,\,j)}_{-}
- \left( \ln \left(y^{(j)}_{-}\right)\right)_x \right) y^{(j)}_{-}
= \frac12 \rho_x^{(k)}
\]
following from relation \rf{bil}.
The above relation gives rise to
\begin{equation}
y^{(j)}_{-\,, x}=- \frac12 \rho_x^{(k)} + 
y^{(j)}_{-} y^{(i,\,j)}_{-}
=  y^{(j)}_{-} \left( y^{(i,\,j)}_{-}- y^{(k)}_{+}\right)+ \alpha_j  \, , \\
\lab{yimx}
\end{equation}
where use was made of equation \rf{bil} in the form
$y^{(j)}_{-} y^{(k)}_{+}= \tfrac12 \rho^{(i)}_{x}$ and where
we introduced 
\[ \alpha_{j} 
= -\frac12 \rho_x^{(k)} + \frac12 \rho_x^{(i)}
=2 \left(v_{i}-v_{k}\right) \, .
\]
Next consider equation \rf{rhoxyi}. It yields :
\[
\rho^{(k)}_x = y^{(k)}_{+,\,x} +y^{(k)}_{+} 
\left(-y^{(k)}_{+} -2 x \right) -2 \nu^{(k)}=
 y^{(k)}_{+,\,x} +y^{(k)}_{+} 
\left(y^{(j)}_{-} + y^{(i,\,j)}_{-} \right) -2 \nu^{(k)}
\]
Rewriting $\rho^{(k)}_x$ as $2 \alpha_{j}   + \rho_x^{(i)}= 
- 2 \alpha_{j} +2 y^{(k)}_{+} y^{(j)}_{-}$ one arrives at 
\begin{equation}
 y^{(k)}_{+,\,x}= y^{(k)}_{+} 
\left(y^{(j)}_{-} -y^{(i,\, j)}_{-} \right) +2 \nu^{(k)}-2  \alpha_{j}=
 y^{(k)}_{+} 
\left(y^{(j)}_{-} -y^{(i,\, j)}_{-} \right) + \alpha_k
\lab{ykpx}
\end{equation}
where for $\nu^{(k)}=\nu =-3 v_{k}$ :
\[
\alpha_k= 2 \nu^{(k)}-2 \alpha_{j}= -6 v_{k} -4 (v_{i}-v_{k})
=2 \left(v_{j}-v_{i}\right)
\]
The final result for the derivative of $y^{(i,\,j)}_{-}$
\begin{equation}
\begin{split}
  y^{(i,\,j)}_{-\,x}&= y^{(i,\, j)}_{-}
\left(y^{(k)}_{+}-y^{(j)}_{-} \right) + \alpha_{i,j}, \\
\alpha_{i,j}&= -\alpha_j-\alpha_k-2= 2 (v_{k}-v_{j})-2
 \end{split}
\lab{yijmx}
\end{equation}
is obtained by taking a derivative of relation 
$y^{(i,\, j)}_{-}+ y^{(j)}_{-}+y^{(k)}_{+}=
-2 x $.
By setting $y^{(k)}_{+}=  f_1, y^{(j)}_{-}=f_2 ,
y^{(i,\, j)}_{-} = f_0$ and $\alpha_k= \alpha_1, \alpha_j=\alpha_2 ,
\alpha_{i,\, j} = \alpha_0$ 
we can summarize the above three equations in a form of the symmetric Painlev\'e
IV equation \cite{noumibk,noumi}:
\begin{equation}
f_{j,x}=f_j\left( f_{j+1}-f_{j+2} \right) +\alpha_j, \;\;
f_{j+3}=f_j,\;\;j=0,1,2
\lab{symmetric}
\end{equation}
Obviously, the association between $f$'s and $y$'s could have 
been chosen differently. This does not matter in view of the 
obvious symmetry for equation \rf{symmetric}:
\begin{equation}
\pi (f_j)=f_{j+1}, \quad \pi (\alpha_j)=\alpha_{j+1}, 
\lab{pisym1}
\end{equation}
\subsection{B\"acklund transformation $\mathbf{G\,:\;y_{-} \to y_{+}}$}
Recall the definition \rf{yrhoi} of  two solutions  $y^{(i)}_{\pm}$ 
of the Painlev\'e IV eq. \rf{painyab} with $b=\nu \pm 1$.
We now define the B\"acklund transformation $G_k$ such 
that:
\begin{equation} G_k (y^{(k)}_{-}) = y^{(k)}_{+}\, .
\lab{Gymk}
\end{equation}
$G_k$ takes a solution of the Painlev\'e IV equation with
$\left( \mu^{(k)}, \nu^{(k)}-2\right)$ to a new solution with 
$\left( \mu^{(k)}, \nu^{(k)}\right)$. As follows from the definition 
\rf{yrhoi} these solutions are related through
\begin{equation}
y^{(k)}_{+}= y^{(k)}_{-}-\left( \ln \left( \rho^{(k)}_{x} \right) \right)_x
\, ,\;\;k=1,2,3.
\lab{ypm}
\end{equation}
Applying $G_k$ on both sides of relation \rf{ypm}
we get 
\begin{equation}
G_k (y^{(k)}_{+}) = y^{(k)}_{+} - G \left( \ln \rho^{(k)}_x\right)_x \,.
\lab{Gypk}
\end{equation}
Applying $G_k$ on both sides of eq. \rf{rhoxyi} we get
$G(\rho^{(k)}_{x})= \rho^{(k)}_{x} -2 y^{(k)}_{+,\, x} -4$,
which after integration yields:
\[
G_k (\rho^{(k)}) = \rho^{(k)} + 2 \left(-2x -y^{(k)}_{+} \right) , 
\]
for all $k=1,2,3$.  Due to the relation \rf{sumsumyik}
we can cast the last equation into
\begin{equation}
G_k (\rho^{(k)}) = \rho^{(k)} + 2 \left(y^{(i,\, j)}_{-}+ y^{(j)}_{-} \right)
\lab{grhok}
\end{equation}
Because $G_k$ increases $\nu=-3v_k$ by $2$ and keeps $\mu^2=
(v_i-v_j)^2$ invariant we find in accordance with the condition 
$\sum_n v_n=0$ and a choice $\mu=v_i-v_j$ that
\begin{equation}
G_k (v_i, v_j, v_k)= \left( v_i-\tfrac13,  v_j-\tfrac13, 
v_k + \tfrac23 \right)
\lab{gk2vvv}
\end{equation}
for all $k$ in agreement with identity 
\rf{Gndiag} given in section \ref{section:intro}.
In terms of the identity \rf{sumsumyik} rewritten as
\begin{equation}
y^{(k)}_{-}+y^{(i)}_{+}+y^{(j,\,i)}_{+} =
y^{(k)}_{+} +y^{(i,\,j)}_{-} +y^{(j)}_{-}=-2x\, ,
\lab{Gidentity}
\end{equation}
$G_k$ maps term by term the three terms on the left hand side of 
the above identity to the three terms the right hand side 
in order of their appearance. 
Applying to identity \rf{Gidentity} the permutation operator $\pi_{ik}$,
which interchanges indices $i$ and $k$ without changing the order of terms 
on both sides, will define $G_i$ as a map which transforms term by term 
the left hand side of such new identity to the right hand side. 
Similarly, applying the permutation operator $\pi_{jk}$ to identity 
\rf{Gidentity} will lead to a new map $G_j$.

We will now introduce three B\"acklund transformations
$g_n$ which square to $G_n$ as given by equations \rf{Gypk}, 
\rf{grhok} and \rf{gk2vvv} for the three different values of $n$
and which generalize relation \rf{ggy}.

We first introduce $g_k$ defined as :
\begin{equation}
g_k \left( y^{(i)}_{+} \right) = y^{(j)}_{-},
\;\;\;
g_k \left( y^{(j)}_{-} \right) = y^{(i,\,j)}_{-} 
=y^{(i)}_{+}+ \left( \ln \left(y^{(j)}_{-}\right)\right)_x
\lab{gykyi} 
\end{equation}
with
\begin{equation}
g_k (y^{(k)}_{+}) = y^{(k)}_{+}- \left( \ln \left(y^{(i,\,j)}_{-}\right)\right)_x
, \lab{gykp} 
\end{equation}
and 
\[
g_k (\rho^{(k)}) = \rho^{(k)}+ 2y^{(j)}_{-}, \;\;
(g_k)^{-1} (\rho^{(k)}) = \rho^{(k)}-2 y^{(i)}_{+}\, .
\]
The choice of indices $i,j,k$ in the above definition adheres with
the identity 
\begin{equation}
y^{(k,\,i)}_{+}+y^{(i)}_{+}+ y^{(j)}_{-}=
y^{(k)}_{+}+y^{(j)}_{-}+ y^{(i,\,j)}_{-} =-2 x  \, ,
\lab{gident}
\end{equation}
which follows from relation \rf{sumsumyik}. 
Since $g_k \left(y^{(k)}_{-}-\left( \ln \left(y^{(i)}_{+}\right)\right)_x 
\right)= y^{(k)}_{+}$ it follows that $g_k$ keeps the above identity 
invariant by mapping term by term the left hand side to the right hand side 
of equation. 
Thus on basis of arguments of subsection \ref{subsection:symmeqs} this observation
shows that $g_k$ is a B\"acklund transformation of the symmetric
Painlev\'e IV equation.
Applying to identity \rf{gident} the permutation operator $\pi_{ik}$,
yields a new map $g_i$ defined explicitly below.
Similarly, applying the permutation operator $\pi_{jk}$ followed by $\pi_{ik}$
gives $g_j$ shown explicitly below.

Furthermore for $g_k$ acting on $y^{(k)}_{+}$ and $ \rho^{(k)}$ it holds 
that $g_k^2=G_k$ as shown  below step-by-step:
\[ \begin{split}
g_k^2 (y^{(k)}_{-})&= g_k^2 \left(y^{(k)}_{+}+ 
\left( \ln \left( \rho^{(k)}_x\right) \right)_x\right)
=  g_k^2 \left(y^{(k)}_{+}+ 
\left( \ln \left( y^{(i)}_{+} y^{(j)}_{-}\right) \right)_x\right)\\
&=g_k \left(y^{(k)}_{+}+ 
\left( \ln \left( y^{(j)}_{-}\right) \right)_x\right)
= y^{(k)}_{+}
\end{split}\]
and
\[ \begin{split}
g_k^2 (y^{(k)}_{+})&=g_k \left(y^{(k)}_{+}- 
\left( \ln \left( y^{(i,\,j)}_{-}\right) \right)_x\right)
=y^{(k)}_{+}- \left( \ln \left( g_k^2
(\rho^{(k)}_x)\right) \right)_x\\
g_k^2 (\rho^{(k)})&= \rho^{(k)}+ 2\left( y^{(j)}_{-}+y^{(i,\,j)}_{-} \right)
= \rho^{(k)}+ 2\left(-2 x - y^{(k)}_{+}\right)
\end{split}\]
for distinct $i,j,k$. The above expressions coincide with the $(g_k)^2=G_k$ 
formula in agreement with \rf{Gymk}, \rf{Gypk} and \rf{grhok}.
{}From first of equations \rf{gykyi}  one sees that $g_k$ maps 
$\nu^{(i)}=-3 v_i$ into $\nu^{(j)}-2=-3 v_j-2$. Thus ${\tilde v}_i=
v_j+\tfrac23$, where ${\tilde v}_i$ is a value of parameter $v_i$ transformed
under $g_k$. {}From identity $(g_k)^2=G_k$  when applied to $ y^{(k)}_{+} $
we find that $g_k$ maps 
$\nu^{(k)}=-3 v_k$ into $\nu^{(j)}+1=-3 v_j+1$ and thus ${\tilde v}_k=
v_k-\tfrac13$. In view of relation $\sum {\tilde v}_n = \sum v_n=0$ we conclude
that 
\begin{equation}
g_k ( v_i,v_j, v_k )\; = \; (v_j +\tfrac23 , v_i-\tfrac13 , v_k -\tfrac13 )\, ,
\lab{gkvvv}
\end{equation}
which shows that $\mu^2=(v_i-v_j)^2\, \to \, (v_i-v_j-1)^2$ under $g_k$.
Acting twice with $g_k$ yields (in agreement with \rf{gk2vvv}) $
G_k ( v_i,v_j, v_k )\; =\; (v_i +\tfrac13 , v_j+\tfrac13 , v_k -\tfrac23 )
$ for which $\mu^2=(v_i-v_j)^2$ remains invariant and $\nu=-3v_k \, \to \, 
\nu+2$.

Furthermore interchanging $i \leftrightarrow k$  
in the above definition of $g_k$ yields $g_i$ defined as :
\begin{equation}
\begin{split}
g_i \left( y^{(j)}_{-} \right)&= y^{(k,\,j)}_{-}=y^{(k)}_{+}
+\left( \ln\left(y^{(j)}_{-}\right)\right)_x\\
g_i \left( y^{(k)}_{+} \right)&=y^{(j)}_{-}\\
g_i (y^{(i)}_{+})&= y^{(i)}_{+}- \left( \ln \left( y^{(k,\,j)}_{-}\right) \right)_x\\
g_i \left(y^{(i,\,j)}_{-}\right)&= y^{(i)}_{+}=
y^{(i,\,j)}_{-}-\left( \ln\left(y^{(j)}_{-}\right)\right)_x\\
g_i \left( \rho^{(i)} \right)&= \rho^{(i)}+2 y^{(j)}_{-}
\end{split}
\lab{gkj}
\end{equation}
Therefore
\[ \begin{split}
g_i^2 (y^{(i)}_{-})&= g_i^2 \left(y^{(i)}_{+}+ 
\left( \ln \left( \rho^{(i)}_x\right) \right)_x\right)
=  g_i^2 \left(y^{(i)}_{+}+ 
\left( \ln \left( y^{(k)}_{+} y^{(j)}_{-}\right) \right)_x\right)\\
&=g_k \left(y^{(i)}_{+}+ 
\left( \ln \left( y^{(j)}_{-}\right) \right)_x\right)
= y^{(i)}_{+}
\end{split}\]
and
\[ \begin{split}
g_i^2 (y^{(i)}_{+})
&= 
y^{(i)}_{+}- \left( \ln \left( g_i^2 (y^{(k)}_{+}) \right) \right)_x-
\left( \ln g_i^2 \left( y^{(j)}_{-}\right) \right)_x\\
g_i^2 (\rho^{(i)})&= \rho^{(i)}+ 2\left( y^{(j)}_{-}+y^{(k,\,j)}_{-} \right)
= \rho^{(i)}+ 2\left(-2 x - y^{(i)}_{+}\right)\, .
\end{split}\]
The above agrees with $g_i^2=G_i$ when acting on $y^{(i)}_{+}, \rho^{(i)}$
variables. Furthermore, from relation \rf{gkvvv} one obtains through
the interchange $i \leftrightarrow k$  :
\begin{equation}
g_i  ( v_i,v_j, v_k )\;=\; (v_i -\tfrac13 , v_k-\tfrac13 , v_j +\tfrac23 )
\lab{givvv}
\end{equation}
and  $ G_i ( v_i,v_j, v_k )\; =\; 
(v_i -\tfrac23 , v_j+\tfrac13 , v_k +\tfrac13 )$. 

We now propose a square root $g_j$ of $G_j$ with respect to 
its action  $G_j \left( y^{(j)}_{-} \right)= y^{(j)}_{+}$.
Define, namely
\begin{equation}
\begin{split}
g_j \left( y^{(j)}_{-} \right)&= y^{(j)}_{-}-\left( \ln\left(y^{(k)}_{+}\right)\right)_x
,\quad
g_j \left(y^{(i,\,j)}_{-}\right)= y^{(k)}_{+} \\
g_j \left( y^{(k)}_{+} \right)&=y^{(i,\,j)}_{-}+\left( \ln\left(y^{(k)}_{+}\right)\right)_x=
y^{(i)}_{-}
\end{split}
\lab{gij}
\end{equation}
Then
\[ \begin{split}
(g_j)^2 (y^{(j)}_{-})&= g_j \left(y^{(j)}_{-}- \left( \ln \left( y^{(k)}_{+}\right) \right)_x\right)\\
&= y^{(j)}_{-}- \left( \ln \left( y^{(k)}_{+} \right) \right)_x-
\left( \ln \left( y^{(i,\,j)}_{-}+\left(\ln \left( y^{(k)}_{+}\right) \right)_x\right) \right)_x\\
&= y^{(j)}_{-}- \left( \ln \left( y^{(k)}_{+} \right) \right)_x-
\left( \ln \left( y^{(i)}_{-}\right) \right)_x = y^{(j)}_{-}- \left( \ln \left( y^{(k)}_{+} y^{(i)}_{-} \right) \right)_x\\
&= y^{(j)}_{-}-\left( \ln \left( \rho^{(j)}_{x} \right) \right)_x= y^{(j)}_{+}
\end{split}\]
Use was made of
\[\begin{split}
y^{(j)}_{+}&= y^{(j)}_{-}-\left( \ln \left( \rho^{(j)}_{x} \right) \right)_x\\
y^{(i)}_{-}&=y^{(i,\,j)}_{-}+\left(\ln \left( y^{(k)}_{+}\right) \right)_x 
\end{split}\]
as follows from definition \rf{yrhoi}.
Comparing with the transformation rules \rf{gij} yields
\begin{equation}
g_j  ( v_i,v_j, v_k )\; = \; (v_k -\tfrac13 , v_j-\tfrac13 , v_i +\tfrac23 )
\lab{gjvvv}
\end{equation}
and $G_j  ( v_i,v_j, v_k )\; =\; 
(v_i +\tfrac13 , v_j-\tfrac23 , v_k +\tfrac13 )$. 

In a compact notation of the symmetric equation \rf{symmetric} based on 
the association $y^{(k)}_{+}=  f_1, y^{(j)}_{-}=f_2 ,
y^{(i,\, j)}_{-} = f_0$  we find :
\begin{xalignat}{3}    
g_k (f_0)&= f_2+\left(\ln f_0\right)_x, & g_i (f_1)&= f_2, &
g_j( f_2)& = f_2 -\left(\ln f_1\right)_x\nonumber\\
g_k (f_1)&= f_1 -\left(\ln f_0\right)_x, & g_i (f_2)&= f_1 +\left(\ln f_2\right)_x, &
g_j( f_0)& = f_1 \lab{gdef1}\\
g_k (f_2)&= f_0, & g_i (f_0)&= f_0-\left(\ln f_2\right)_x, &
g_j( f_1)& = f_0 +\left(\ln f_1\right)_x \, .\nonumber
\end{xalignat}
The relevant algebraic relations for the B\"acklund transformations are
given in \rf{grelations}. In addition it holds that 
\begin{equation}
g_k\, g_i = g_i\, g_j= g_j\,g_k= \pi^2 \, ,
\lab{pig}
\end{equation}
where the permutation operator $\pi$ has been defined in \rf{pisym1} 
and based on the above relations satisfies $ 
\pi\, ( v_i,v_j, v_k )= 
(v_k -\tfrac13 , v_i-\tfrac13 , v_j+\tfrac23 )$ as is consistent
with $\pi^3=1$. 
Relation \rf{pig} together with relations \rf{grelations} 
provide realization of the affine Weyl group symmetry in terms of
three B\"acklund transformations $g_n, n=1,2,3$.
Note that the last two identities of relations \rf{grelations} amount to
\[
G_j^{-1}=g_k G_j g_k, \quad G_i^{-1}=g_j G_i g_j, \quad G_k^{-1}=g_i G_k
g_i\, .
\]
Combining some of the above operations one finds:
\[
\pi G_i=G_k \pi, \quad \pi G_k=G_j \pi, \quad
 \pi G_j=G_i \pi,
\]
and
\begin{equation}
g_j= G_i^{-1} \pi_{ik}, \quad g_i= G_j^{-1} \pi_{jk}, \quad 
g_k= G_j^{-1} \pi_{ij} \, .
\lab{gnewdef}
\end{equation}
Equation \rf{gnewdef} shows that a composition of the DB and the
permutation transformations yields the B\"acklund transformations
$g_n, n=1,2,3$. 
This observation will come to good use in section \ref{section:applications} as a tool
to construct a set of Painlev\'e IV solutions which closes under the B\"acklund transformations.

We now write down three transformations :
\begin{equation} 
 s_0=g_k \,\pi^2, \;\;s_1=g_j \pi^2 , 
\;\;s_2=g_i\,\pi^2
\lab{ses}
\end{equation}
introduced in \cite{noumi} (and pedagogicaly described in 
\cite{noumibk}) to conveniently interpret the above three 
maps in terms of root systems associated to the affine Weyl group $A_2$.

It now follows from relations \rf{gdef1} that these three transformations satisfy
\begin{xalignat}{3}    
s_0 (f_0)&= f_0, & s_1 (f_1)&= f_1, &
s_2( f_2)& = f_2 \nonumber\\
s_0 (f_1)&= f_2 +\left(\ln f_0\right)_x, & s_1 (f_2)&= f_0 +\left(\ln f_1\right)_x, &
s_2( f_0)& = f_1 +\left(\ln f_2\right)_x \lab{sdef1}\\
s_0 (f_2)&= f_1-\left(\ln f_0\right)_x, & s_1 (f_0)&= f_2-\left(\ln f_1\right)_x, &
s_2( f_1)& = f_0 -\left(\ln f_2\right)_x \nonumber
\end{xalignat}
and 
\begin{equation}
\begin{split}
s_0\,(v_i,v_j, v_k )\; &=\; 
(v_i , v_k-1, v_j +1 ) \\
s_1\,(v_i,v_j, v_k )\; &=\; 
(v_j , v_i, v_k ) \\
s_2\,(v_i,v_j, v_k )\; &=\; 
(v_k , v_j, v_i ) 
\lab{pisvis}
\end{split}
\end{equation}
One easily finds that $s_i^2=1, i=0,1,2$. 

Based on definition \rf{sdef1} one checks that $s_1 (y^{(j)}_{-})
=y^{(i)}_{-}, s_1 (y^{(i)}_{-}) =y^{(j)}_{-}$ and 
$s_2 (y^{(i)}_{+})
=y^{(k)}_{+}, s_2 (y^{(k)}_{+}) =y^{(i)}_{+}$.
Thus
\begin{equation}
s_1= \pi_{ij},\quad s_2= \pi_{ik}
\lab{spik}
\end{equation}
as already suggested by relation \rf{pisvis}.

The following ``inverse'' relations :
\[ g_k =s_0 \pi = \pi s_2 ,\;\; g_i=s_2 \pi = \pi s_1, \;\;
g_j=s_1 \pi = \pi s_0 
\]
hold between $s$- and $g$-B\"acklund transformations.

Let us consider the $s_1$ and $s_2$ maps from eq. \rf{pisvis}
and set for simplicity $(j,i, k)=(1,2, 3)$. Then $s_1$ and $s_2$ maps
can be geometrically realized as reflections in the hyperplane 
orthogonal to vectors $\vec{\alpha}_1=\vec{e}_1-\vec{e}_2$ and $
\vec{\alpha}_2=\vec{e}_2-\vec{e}_3$. The $s_0$ map from eq. \rf{pisvis}
can then be realized as a reflection in a hyperplane $\{\vec{x}: \langle 
\vec{\alpha}_0 \vert \vec{e}_j-\vec{e}_k\rangle=1 \}$, with
$\vec{\alpha}_0=\vec{e}_1-\vec{e}_3$ being the highest root of $A_2$
(see f.i. \cite{forrester}).
In terms of $\alpha_{j} =\alpha_2 =2 \left(v_{2}-v_{3}\right)$,
$\alpha_k= \alpha_1=2 \left(v_{1}-v_{2}\right)$ and
$\alpha_{i,j}=\alpha_0= -\alpha_j-\alpha_k-2= 2 (v_{3}-v_{1})-2$,
introduced in subsection \ref{subsection:symmeqs},
the $s_i$'s maps can be expressed by the Cartan matrix $A_{ij}$
\cite{noumibk}:
\[
\alpha_j \; \stackrel{s_i}{\longrightarrow}\; 
\alpha_j- A_{ij} \alpha_i, \quad i,j=0,1,2
\]

Thus we recover from our symmetry structure the extended 
affine Weyl group structure 
in the form denoted by Noumi in \cite{noumibk} as ${\widetilde W}= \langle s_0,s_1,s_2;\pi\rangle$.
It is here introduced only on basis of the three B\"acklund transformations
$g_n, n=1,2,3$. 

\subsection{From symmetric Painlev\'e equation to equation for $\rho$}
We now take as a starting point the symmetric Painlev\'e equation \rf{symmetric}.
Our goal is to derive the $\rho$-function and its equation \rf{jmo2}
solely from the quantities entering  equation \rf{symmetric}.
Connection from the symmetric Painlev\'e IV equation to the Hamiltonian
formalism was already established in \cite{noumiya}. This subsection
provides an alternative proof for that relation.

Out of the three transformation $g_n ,n=1,2,3$ we will chose
$g_j$ defined in eq. \rf{gij} to illustrate our construction. 
We denote the action of $g_j$  by $\,{\tilde{}}\,$. Then
\begin{xalignat}{2}
\tilde{f_2} &= f_{2}-\left( \ln f_{1}\right)_x= f_{0} - \frac{\alpha_{1}}{f_{1}}, &
   \tilde{\alpha}_{2}&= -2-\alpha_{2} \nonumber  \\
\tilde{f_0}&=f_{1},&\tilde{\alpha}_{0}&= -\alpha_{1} \lab{tildatrans} \\
\tilde{f_1}&= f_{0}+\left( \ln f_{1}\right)_x= f_{2} + \frac{\alpha_{1}}{f_{1}}
,&\tilde{\alpha}_{1}&= \alpha_{1}+\alpha_{2} \nonumber
\end{xalignat}
For the transformed functions it still holds
that $\tilde{f_0}+ \tilde{f_1}+ \tilde{f_2}=-2x$.
Also they satisfy  the symmetric Painlev\'e
eq. \rf{symmetric} with the coefficients $\tilde{\alpha}_{i}, i=0,1,2$ given
above.

Acting twice with this transformation yields
\begin{xalignat}{2} 
{\bar f_0}&= {\tilde{\tilde{f_0}}}= {\tilde f_1}=f_2 +\frac{\alpha_1}{f_1}&
 \bar{\alpha}_0&=-\alpha_{1}-\alpha_2  \nonumber\\
{\bar f_2}&= {\tilde{\tilde{f_2}}}= f_1
  \frac{f_1f_2-\alpha_2}{f_1f_2+\alpha_1}&
 \bar{\alpha}_2&= \alpha_2, \lab{bartransf}\\
{\bar f_1}&= {\tilde{\tilde{f_1}}}= f_0-\frac{\alpha_1}{f_1}+\frac{\alpha_1+\alpha_2}{f_2+\alpha_1/f_1}&
\bar{\alpha}_1&= -2 +\alpha_1 \nonumber
\end{xalignat} 
which describes an actio‭n of $G_j$. Because $G_{j}$ operation is a symmetry
it follows that ${\bar f_i}, i=0,1,2$ satisfy as well the symmetric Painlev\'e
eq. \rf{symmetric} with coefficients $\bar{\alpha_{i}}, i=0,1,2$ given
above.
Next, define the quantities :
\begin{equation}\begin{split}
\frac12 \sigma^{(j)}_x&=f_0f_1+f_{1\, x}= f_1f_2+\alpha_1,\quad
\frac12 \sigma^{(i)}_x=f_1f_2,\\
\frac12 \sigma^{(k)}_x&=f_0f_2-f_{2\, x}= f_1f_2-\alpha_2
\end{split}
\lab{rhosj}
\end{equation}
It follows that
\begin{equation}
\bar{f_2} - f_{2} =- \left(\ln (f_{1}f_{2}+\alpha_{1})\right)_{x}
= - \left(\ln \sigma^{(j)}_x\right)_{x}
\lab{fbarf2}
\end{equation}
Also the following important Riccati equation 
\begin{align}
&-f_{2,\,x}-f_2^{2}-2 x f_2 = 
-f_2\left( f_{0}-f_1 \right) -\alpha_2-f_2^{2} +(f_0+f_1+f_2)f_2 \nonumber\\
&=  2 f_1 f_{2}-\alpha_{2}=\sigma^{(j)}_x
-2 \alpha_{1} -\alpha_{2}
\lab{riccat1}
\end{align}
follows from the symmetric Painlev\'e
eq. \rf{symmetric}.

We want to show that the transformed version of  ${\sigma}^{(j)}_x$ given by
\[\frac12 {\bar \sigma}^{(j)}_x ={\bar f}_0 {\bar f}_1+{\bar f}_{1\, x}
\]
is equal to
\[\frac12 {\bar \sigma}^{(j)}_x =
\frac12 \sigma^{(j)}_x -2 - {\bar f}_{2\, x}
\]
We accomplish this in two steps. First we apply the ${}^{\tilde{}}$
transformation on relation \rf{rhosj} which yields
\[
\frac12 \tilde{\sigma}^{(j)}_x =\tilde{f}_0 \tilde{f}_1+\tilde{f}_{1\, x}
=f_1 (f_{0}+( \ln f_{1})_x) +\tilde{f}_{1\, x}= f_0f_1 +  f_{1, x} + \tilde{f}_{1\,x}=
\frac12 {\sigma}^{(j)}_x +\tilde{f}_{1\,x}
\]
Applying another ${}^{\tilde{}}$ transformation on
the above equation yields
\[
\begin{split}
\frac12 \bar{\sigma}^{(j)}_x &=
\frac12 \tilde{\sigma}^{(j)}_x+\bar{f}_{1\, x}
=\frac12 {\sigma}^{(j)}_x +\tilde{f}_{1,x}+\bar{f}_{1\, x}\\
&=\frac12 {\sigma}^{(j)}_x +\bar{f}_2 (\bar{f}_1-\bar{f}_0)-2 -{\alpha}_2
=\frac12 {\sigma}^{(j)}_x -2 -\bar{f}_{2 \,x}
\end{split}
\]
where we used : $\tilde{f}_0= \bar{f}_2+\tilde{\alpha}_1/ \bar{f}_0$,
$  \tilde{f}_1=\bar{f}_0$ and $\tilde{f}_2=\bar{f}_1-\tilde{\alpha}_1/ \bar{f}_0$
and $\bar{\alpha}_1 - \tilde{\alpha}_1=  -2 -{\alpha}_2$.

Repeating steps taken in the derivation of the Riccati equation \rf{riccat1}
and using the result we have established above we arrive at
\begin{equation}
-\bar{f}_{2,\,x}-\bar{f}_2^{2}-2 x \bar{f}_2 =  \bar{\sigma}^{(j)}_x
-2 \bar{\alpha}_{1} -\bar{\alpha}_{2}\nonumber
=\sigma^{(j)}_x-2 \bar{f}_{2\,x} -2 {\alpha}_{1} -{\alpha}_{2}  
\lab{riccat2}
\end{equation}
Thus we have established a Darboux dressing chain relation for $f_{2}$
and $\bar{f}_2$  :
\begin{equation}
-f_{2,\,x}-f_2^{2}-2 x f_2 =
\bar{f}_{2,\,x}-\bar{f}_2^{2}-2 x \bar{f}_2 
=\sigma^{(j)}_x
-2 \alpha_{1} -\alpha_{2}\, .
\lab{dchain}
\end{equation}
Plugging on the right hand side of the above equation the expression 
\rf{fbarf2} for $\bar{f}_2$ in terms of $f_{2}$ and subtracting 
from the left hand side yields
\begin{equation}
f_{2,\,x}+f_2 \left( \ln \sigma^{(j)}_x\right)_{x}= \frac12  \left( \ln \sigma^{(j)}_x\right)_{xx}
+ \frac12  \left( \ln \sigma^{(j)}_x\right)_{x}^{2} -x  \left( \ln \sigma^{(j)}_x\right)_{x}
\lab{firstorder}
\end{equation}
This is a first order differential equation for which we easily find
a particular solution
\begin{equation}
f_{2} = \frac{-1}{2\sigma^{(j)}_x}\left(2 \left(x
  \sigma^{(j)}_x-\sigma^{(j)}\right)-\sigma^{(j)}_{xx}\right)
\lab{f2rho}
\end{equation}
Note that adding to the above a homogeneous solution of equation 
\rf{firstorder} amounts to adding an integration constant
to $\sigma^{(j)}$.

Furthermore from the relation \rf{fbarf2} one gets
\begin{equation}
\bar{f}_{2} =  \frac{-1}{2\sigma^{(j)}_x}\left(2 \left(x
  \sigma^{(j)}_x-\sigma^{(j)}\right)+\sigma^{(j)}_{xx}\right)
\lab{barf2rho}
\end{equation}
The last two equations agree with relation \rf{yrhok}
for $\sigma^{(j)}=\rho^{(j)}$, $f_2=y^{(j)}_{-}$ and 
$\bar{f}_{2}=y^{(j)}_{+}$. 

Recall from \rf{bartransf} that
\[
\bar{f}_{2} =f_{1} \frac{f_{1}f_{2}-\alpha_{2}}{f_{1}f_{2}+\alpha_{1}}
\]
{}from the above relation and definition \rf{rhosj} we conclude that
\begin{equation}
f_{2 }\bar{f}_{2}= f_{1}f_{2 } \frac{f_{1}f_{2}-\alpha_{2}}{f_{1}f_{2}+\alpha_{1}} 
=\frac12 \sigma^{(i)}_{x} \, \frac12 \sigma^{(k)}_{x}\, \left(\frac12
\sigma^{(j)}_{x}\right)^{-1}  
\lab{f2rhoa}
\end{equation}
Plugging into the above expressions \rf{f2rho} and \rf{barf2rho}
we recover
\begin{equation}
\left(2 \left(x
  \sigma^{(j)}_x-\sigma^{(j)}\right)-\sigma^{(j)}_{xx}\right)
\left(2 \left(x
  \sigma^{(j)}_x-\sigma^{(j)}\right) +\sigma^{(j)}_{xx}\right)=
 2 \sigma^{(i)}_{x} \, 
\sigma^{(j)}_{x}\,  \sigma^{(k)}_{x}
\lab{rhojeq}
\end{equation}
which shows that $\sigma^{(j)}$ satisfies the $\rho$-equation \rf{jmo2}.
Since the left hand side of equation \rf{rhojeq} is unaffected 
by a shift of $\sigma_x$ by a constant the relation also holds for $\sigma^{(i)}$
and $\sigma^{(k)}$. The same conclusion also follows from the same 
consideration as given above when applied to $g_{i}$ and $g_{k}$.

\section{Applications}
\label{section:applications}
\subsection{the ``$-2x$ and $-1/x$-hierarchies}
Recall a basic solution of the ``$-2x$ hierarchy''
\cite{Clarkson-jmp,Clarkson,kajiwara,noumi}
\begin{equation}
\rho_{(k,n)} = 2 \partial_x \ln W_{k} \lbrack H_n, H_{n-1}, {\ldots} ,
H_{n-k+1}\rbrack \,, \;\;\; k>1\, ,
\lab{rhokn}
\end{equation}
which is  directly obtained as a reduction of the soliton solutions
of the AKNS hierarchy as discussed in \cite{AGZ2009}. 
Here, $H_n(x)$ is the Hermite polynomial and $\rho_{(k,n)}$
satisfies $\rho$-equation \rf{jmo2} with $\nu
=n-2 k+1$ and $\mu^2=(n+1)^2$. 
Formula \rf{yrho} (with a minus subscript) yields:
\begin{equation}
y_{-, (k,n)} = 
\partial_x \ln \frac{W_{k+1} \lbrack {H}_n, { H}_{n-1},{\ldots} , 
{ H}_{n-k}\rbrack}{W_{k} \lbrack {H}_n, { H}_{n-1},{\ldots} , 
{ H}_{n-k+1}\rbrack} -2 x \, ,
\lab{wyfr}
\end{equation}
which solves the Painlev\'e IV equation with the same values of $\nu$ and $\mu$
as the solution in \rf{rhokn}.

Next, we implement $\rho \to \rho^{(i)}, i=1,2$
symmetry to derive from the above other solutions.
We proceed as in \rf{kone}-\rf{ktwo} with $\mu=(n+1)$ and
$\nu =n-2\cdot k+1$ :
\begin{align}    
\rho^{(1)}_{(k,n)}&= \rho_{(k,n)}-2\cdot 2kx,  \;\;\nu^{(1)}
= n+k+1, \;\; \mu^{(1)} = \pm (n-k+1) 
\lab{kone1}\\
\rho^{(2)}_{(k,n)}&= \rho_{(k,n)}+2 \cdot 2 (n-k+1)x,  \;\;\nu^{(2)}
= - 2n+k-2,\; \; \mu^{(2)} = \pm k \lab{ktwo1}
\end{align}
Thus
\begin{equation}
\rho^{(1)}_{(k,n)}
= 2 \partial_x \ln W_{k} \lbrack e^{-x^2} H_n, e^{-x^2} H_{n-1}, {\ldots} ,
e^{-x^2} H_{n-k+1}\rbrack\, , \lab{rho1x}
\end{equation}
which via formula \rf{yrho}(with the ``$+$'' subscript)  gives rise to 
\begin{equation}
y^{(1)}_{(k,n)}
=  \partial_x \ln \frac{W_{k} \lbrack {H}_{n}, { H}_{n-1},
{\ldots} , { H}_{n-k+1}\rbrack}{W_{k+1} \lbrack {H}_{n+1}, { H}_{n},{\ldots} , 
{ H}_{n-k+1}\rbrack}
=\partial_x \ln \frac{W_{n-k+1} \lbrack {\widehat H}_{n}, {\widehat H}_{n-1},{\ldots} , 
{\widehat H}_{k}\rbrack}{W_{n-k+1} \lbrack {\widehat H}_{n+1}, {\widehat H}_{n},
{\ldots} , {\widehat H}_{k+1}\rbrack}\, ,
\lab{y1kn}
\end{equation}
where ${\widehat H}_m (x) = e^{-x^2} d^m e^{x^2}/d x^m
= (- \iu)^m H_m (\iu x)$.
The function in eq. \rf{y1kn} solves the Painlev\'e IV equation with 
$\nu= n+k+1$ and $\mu^2= (n-k+1)^2$ and belongs to the so-called 
``$-1/x$-hierarchy'' but here is obtained as a result of the simple
symmetry transformation $\rho \to \rho^{(1)}$.

A convenient expression for $\rho^{(2)}_{(k,n)}$
from equation \rf{ktwo1} is
\begin{equation}
\rho^{(2)}_{(k,n)}= 2 \partial_x \ln W_{n-k+1} \lbrack e^{x^2} {\widehat H}_n, 
e^{x^2} {\widehat H}_{n-1}, {\ldots} ,e^{x^2} {\widehat H}_{k}\rbrack 
\lab{rho2x}
\end{equation}
and via formula \rf{yrho} (with ``$+$'' subscript)  it gives rise to 
\begin{equation}
y^{(2)}_{(k,n)}
=  \partial_x \ln \frac{W_{n-k+1} \lbrack {\widehat H}_{n}, {\widehat H}_{n-1},
{\ldots} , {\widehat H}_{k}\rbrack}{W_{n-k} \lbrack {\widehat H}_{n-1}, {\ldots} , 
{\widehat H}_{k}\rbrack}
= \partial_x \ln \frac{W_{k} \lbrack {H}_{n}, {H}_{n-1},
{\ldots} , { H}_{n-k+1}\rbrack}{W_{k} \lbrack {H}_{n-1}, {H}_{n-2},{\ldots} , 
{H}_{n-k}\rbrack}
\lab{y2kn}
\end{equation}
which solves the Painlev\'e IV equation with 
$\nu= -2n+k-2$ and $\mu^2= k^2$. 
Substituting $k^{\prime}=n-k, n^{\prime}=n-1$ we can rewrite the above
as:
\begin{equation}
y^{(2)}_{(k,n)}=  \partial_x \ln \frac{W_{k^{\prime}+1} \lbrack {\widehat H}_{n^{\prime}+1}, { \widehat H}_{n^{\prime}},{\ldots} , 
{ \widehat H}_{n^{\prime}-k^{\prime}+1}\rbrack}{W_{k^{\prime}} \lbrack {\widehat H}_{n^{\prime}}, { \widehat H}_{n^{\prime}-1},
{\ldots} , { \widehat H}_{n^{\prime}-k^{\prime}+1}\rbrack}
= \partial_x \ln \frac{W_{n^{\prime}-k^{\prime}+1} \lbrack {H}_{n^{\prime}+1}, {H}_{n^{\prime}},
{\ldots} , { H}_{k^{\prime}+1}\rbrack}{W_{n^{\prime}-k^{\prime}+1} \lbrack {H}_{n^{\prime}}, {H}_{n^{\prime}-1},{\ldots} , 
{H}_{k^{\prime}}\rbrack}
\lab{hwiiiq}
\end{equation}
Also $y^{(2)}_{(k,n)}$ denoted as $w^{(k^{\prime},n^{\prime})}$ 
belongs to the so-called ``$-1/x$-hierarchy'' with the
parameters 
$\nu=-n^{\prime}-k^{\prime}-3, \mu= \pm \left( n^{\prime}-k^{\prime}+1\right)$.

The following version of identity \rf{sumsumyik} holds here :
\begin{equation}
y_{-, (k,n)}+ y^{(1)}_{(k,n)}+y^{(2)}_{(k,n)}= -2 x-
\left( \ln y_{-, (k,n)}\right)_x
\lab{wkp1}
\end{equation}

Next, recall (c.f. \cite{AGZ2009}) the remaining basic solution of 
the ``$-2x$ hierarchy''
\[
{\widehat \rho}^{(k,n)} = 2 \partial_x \ln W_{k} \lbrack {\widehat H}_n, {\widehat H}_{n-1},{\ldots} , 
{\widehat H}_{n-k+1}\rbrack
\]
which satisfies $\rho$-equation \rf{jmo2} with $\nu
=-n+2\cdot k-1$ and $\mu^2=(n+1)^2$.
Using relation \rf{yrho} (with a ``minus'' subscript) yields the corresponding
solution to the Painlev\'e IV equation :
\begin{equation}
{\widehat y}^{(k,n)}_{-} =  -\partial_x \ln \frac{W_{k} \lbrack {\widehat H}_n, {\widehat H}_{n-1},{\ldots} , 
{\widehat H}_{n-k+1}\rbrack}{W_{k-1} \lbrack {\widehat H}_n, {\widehat H}_{n-1},{\ldots} , 
{\widehat H}_{n-k+2}\rbrack} -2 x\\
\lab{wyfq}
\end{equation}
with $b= -n+2k-2=\nu-1$ and $\mu^2=(n+1)^2$.

We proceed as in \rf{kone}-\rf{ktwo} with $\mu=(n+1)$ to obtain :
\begin{align}    
{\widehat \rho}^{(1)}_{(k,n)}&={\widehat \rho}_{(k,n)}-2 \cdot 2 (n-k+1)x
,  \;\;\nu^{(1)}
= 2n-k+2, \;\; \mu^{(1)} = \pm k \nonumber\\
&= 2 \partial_x \ln W_{n-k+1} \lbrack e^{-x^2} {H}_n, 
e^{-x^2} {H}_{n-1}, {\ldots} ,e^{-x^2} {H}_{k}\rbrack 
\lab{hkone1}\\
{\widehat \rho}^{(2)}_{(k,n)}&={\widehat  \rho}_{(k,n)}+2\cdot 2kx,
  \;\;\nu^{(2)}
= -n-k-1,\; \; \mu^{(2)} = \pm (n-k+1)\nonumber \\
&= 2 \partial_x \ln W_{k} \lbrack e^{x^2} {\widehat H}_n, 
e^{x^2} {\widehat H}_{n-1}, {\ldots} ,e^{x^2} {\widehat H}_{n-k+1}\rbrack 
\lab{hktwo1}
\end{align}
Via formula \rf{yrho} they give rise to 
\begin{equation}
{\widehat y}^{(1)}_{(k,n)}
=  \partial_x \ln \frac{W_{n-k+2} \lbrack {H}_{n+1}, {H}_{n-1},
{\ldots} , {H}_{k}\rbrack}{W_{n-k+1} \lbrack {\widehat H}_{n}, {\ldots} , 
{H}_{k}\rbrack}
= \partial_x \ln \frac{W_{k} \lbrack {\widehat H}_{n+1}, {\widehat H}_{n-1},
{\ldots} , {\widehat  H}_{n-k+2}\rbrack}{W_{k} \lbrack {\widehat H}_{n}, {\widehat H}_{n-2},{\ldots} , 
{\widehat H}_{n-k+1}\rbrack}
\lab{hy2kn}
\end{equation}
which solves the Painlev\'e IV equation with 
$\nu=2n-k+2$ and $\mu^2= k^2$ and
\begin{equation}
{\widehat y}^{(2)}_{(k,n)}
=  \partial_x \ln \frac{W_{k} \lbrack {\widehat H}_{n}, {\widehat H}_{n-1},
{\ldots} , {\widehat H}_{n-k+1}\rbrack}{W_{k-1} \lbrack {\widehat H}_{n-1}, {\ldots} , 
{\widehat H}_{n-k+1}\rbrack}
= \partial_x \ln \frac{W_{n-k+1} \lbrack {H}_{n}, {H}_{n-1},
 {\ldots} , { H}_{k}\rbrack}{W_{n-k+1} \lbrack {H}_{n-1}, {H}_{n-2},{\ldots} , 
 {H}_{k-1}\rbrack}
\lab{hy2kn1}
\end{equation}
with $\nu^{(2)}
= -n-k-1,\; \; \mu^{(2)} = \pm (n-k+1)$.

It holds that
\begin{equation}
{\widehat y}^{(k,n)}_{-}+ {\widehat y}^{(1)}_{(k,n)}+ {\widehat y}^{(2)}_{(k,n)}
=-2x - \left(\ln {\widehat y}^{(k,n)}_{-}\right)_x \, .
\end{equation}

\subsection{ ``$-2x/3$-hierarchy''}
We will introduce the following polynomials :
\begin{equation}
F_n^{(k)}=  \frac{e^{x^2/3}}{2^n n!} \dder[3n+k]{}{x} e^{-x^2/3},\;\;
{\widehat F}_n^{(k)}= \, \frac{ e^{-x^2/3}}{2^n n!} \dder[3n+k]{}{x} e^{x^2/3},
\lab{fnk}
\end{equation}
defined for  $k=0,1,2, \;\;n=0,1,2,{\ldots}$.

A starting point is a basic polynomial solution of 
equation \rf{jmo2}:
\begin{equation}
\rho^{(0)} = \frac{8}{27} x^3 , \quad (\mu^{(0)})^2 = \frac49,
\;\; \nu^{(0)}= 0\, .
\lab{rhob1}
\end{equation}
We proceed as in relations \rf{kone}-\rf{ktwo} with $\mu^{(0)}=2/3$
to obtain :
\begin{align}
\rho^{(1)} &= \frac{8}{27} x^3 - \frac{4}{3} x, \quad (\mu^{(1)})^2 
= \frac19,\;\; \nu^{(1)} = + 1\,
\lab{rho13m}\\
\rho^{(2)} &= \frac{8}{27} x^3 + \frac{4}{3} x, \quad (\mu^{(2)})^2  
= \frac19, \;\; \nu^{(2)}  = - 1\,.
\lab{rho13p}
\end{align}
Then repeated actions with $G^{(\pm 1)}$ on $\rho^{(0)}$ yield:
\begin{equation}
\begin{split}
\rho^{(0,n)}&=G^{n} (\rho^{(0)})=
\rho^{(0)} - 2n \frac{4x}{3}+  
2 \left(\ln W_{n} \lbrack F_0^{(1)} , F_1^{(1)}, {\ldots} , 
F_{n-1}^{(1)}\rbrack\right)_x\\
&=\rho^{(0)}
+  2 \left(\ln W_{n} \lbrack e^{\frac{-2x^2}{3}} F_0^{(1)} ,
e^{\frac{-2x^2}{3}} F_1^{(1)}, {\ldots} , e^{\frac{-2x^2}{3}}
F_{n-1}^{(1)}\rbrack\right)_x
\;\; \nu=2n \\
\rho^{(0,-n)}&=G^{-n} (\rho^{(0)})=
\rho^{(0)} + 2n \frac{4x}{3}+  
2 \left(\ln W_{n} \lbrack {\widehat F}_0^{(1)} , {\widehat F}_1^{(1)}, {\ldots} , 
{\widehat F}_{n-1}^{(1)}\rbrack\right)_x\\
&=\rho^{(0)} +2 \left(\ln W_{n} \lbrack 
e^{\frac{2x^2}{3}} {\widehat F}_0^{(1)} , 
e^{\frac{2x^2}{3}} {\widehat F}_1^{(1)}, {\ldots} , e^{\frac{2x^2}{3}}
{\widehat F}_{n-1}^{(1)}\rbrack\right)_x
\;\; \nu=-2n \, ,
\end{split}
\lab{rhowrp}
\end{equation}
which are solutions of the $\rho$-equation \rf{jmo2}
with $\mu^2 = \frac49$ and $\nu=\pm 2n, n=1,2, 3, {\ldots}$, 
respectively.

Next, proceeding as in \rf{kone}-\rf{ktwo} with respect to
$\rho^{(0,n)}$ yields from \rf{rhowrp} :
\begin{align}
\rho^{(1,n)} &= \rho^{(1)} +n \frac{4x}{3}+  
2 \left(\ln W_{n} \lbrack F_0^{(1)} , F_1^{(1)}, {\ldots} , 
F_{n-1}^{(1)}\rbrack\right)_x, \nonumber \\
\mu^{(1,n)}&= \pm \left(\frac13+n\right),\;\; 
\nu^{(1,n)} = 1-n \lab{rho1n}\\
\rho^{(2,n)} &= \rho^{(2)} +n \frac{4x}{3}+  
2 \left(\ln W_{n} \lbrack F_0^{(1)} , F_1^{(1)}, {\ldots} , 
F_{n-1}^{(1)}\rbrack\right)_x, \nonumber \\
\mu^{(2,n)}  &= \pm\left(\frac13-n\right), \;\; \nu^{(2,n)}  = - 1-n\,.
\lab{rho2n}
\end{align}
Acting repeatedly with $G$ according to
\[
G (\rho^{(i,n)}) = \rho^{(i,n)} + 2 \left(-2x -y^{(i,n)}_{+} \right),
\quad i=1,2
\]
yields after $k$-times:
\begin{align}
\rho^{(1,n,k)} &= \rho^{(1)} +\left(n -2k\right)\frac{4x}{3} +  
2 \left(\ln W_{n} \lbrack F_0^{(1)} , F_1^{(1)}, {\ldots} , 
F_{n-1}^{(1)},  F_0^{(2)}, F_1^{(2)}, {\ldots} , F_{k-1}^{(2)} \rbrack\right)_x, \nonumber \\
\mu^{(1,n,k)}&= \pm \left(\frac13+n\right),\;\; 
\nu^{(1,n,k)} = 1-n+2k \lab{rho1nk}\\
\rho^{(2,n,k)} &= \rho^{(2)} +\left(n -2k\right) \frac{4x}{3}+  
2 \left(\ln W_{n} \lbrack F_0^{(1)} , F_1^{(1)}, {\ldots} , 
F_{n-1}^{(1)},F_0^{(0)}, F_1^{(0)}, {\ldots} , F_{k-1}^{(0)}
\rbrack\right)_x, \nonumber \\
\mu^{(2,n,k)}  &= \pm\left(\frac13-n\right), \;\; \nu^{(2,n,k)}  = - 1-n+2k\,.
\lab{rho2nk}
\end{align}
The same steps, meaning, first applying \rf{kone}-\rf{ktwo} with respect to
$\rho^{(0,-n)}$ from \rf{rhowrp} in order to derive
$\rho^{(i,-n)}, i=1,2$ and then applying $G^{-1}$ gives : 
\begin{align}
\rho^{(1,-n,-k)} &= \rho^{(1)} -\left(n -2k\right)\frac{4x}{3} +  
2 \left(\ln W_{n} \lbrack {\widehat F}_0^{(1)} , {\widehat F}_1^{(1)}, {\ldots} , 
{\widehat F}_{n-1}^{(1)},  {\widehat F}_0^{(0)}, {\widehat F}_1^{(0)}, {\ldots} , 
{\widehat F}_{k-1}^{(0)} \rbrack\right)_x, \nonumber \\
\mu^{(1,-n,-k)}&= \pm \left(\frac13-n\right),\;\; 
\nu^{(1,-n,-k)} = 1+n-2k \lab{rho1mnk}\\
\rho^{(2,-n,-k)} &= \rho^{(2)} -\left(n -2k\right) \frac{4x}{3}+  
2 \left(\ln W_{n} \lbrack {\widehat F}_0^{(1)} , {\widehat F}_1^{(1)}, {\ldots} , 
{\widehat F}_{n-1}^{(1)}, {\widehat F}_0^{(2)}, {\widehat F}_1^{(2)}, {\ldots} , 
{\widehat F}_{k-1}^{(2)}
\rbrack\right)_x, \nonumber \\
\mu^{(2,-n,-k)}  &= \pm\left(\frac13+n\right), \;\; \nu^{(2,-n,-k)}  = - 1+n-2k\,.
\lab{rho2mnk}
\end{align}
One finds by comparing $\mu$ and $\nu$'s of the above expressions that 
\[
\rho^{(2,n, k)}= \rho^{(1, -n, -(1+n-k))},  \quad \, n,k =0,1,2,3, {\ldots} , \;\;
\]
\[
\rho^{(1,n,k)}= \rho^{(2,-n, -(n-1-k))},  \quad n,k=0,1,2,3, {\ldots} , \;\;
\]
Equation \rf{yrhoi} is providing a general construction for
a map $\rho^{(i,n, k)} \to y^{(i,n, k)}_{+}$ and 
$\rho^{(i,-n, -k)} \to y^{(i,-n, -k)}_{-}$ for $i=1,2$ with
the following results:
\begin{align}
y^{(1,n, k)}_{+}&=-\left(\ln \left(
\frac{W_{k+n+1} \bigl\lbrack 
F_0^{(1)}, F_1^{(1)}, {\ldots} , F_{n-1}^{(1)},
F_0^{(2)}, F_1^{(2)}, {\ldots} , F_k^{(2)}
\bigr\rbrack}
{W_{k+n} \bigl\lbrack F_0^{(1)}, F_1^{(1)}, {\ldots} , F_{n-1}^{(1)},
F_0^{(2)}, F_1^{(2)}, {\ldots} , F_{k-1}^{(2)}
\bigr\rbrack}\right)\right)_x-\frac{2x}{3}, \lab{y1nk}\\
\mu^2&=\left(\frac{1}{3}+n\right)^2 ,\qquad \nu=2k-n+1  \nonumber\\
y^{(2,n, k)}_{+}&=-\left(\ln \left(
\frac{W_{k+n+1} \bigl\lbrack 
F_0^{(1)}, F_1^{(1)}, {\ldots} , F_{n-1}^{(1)},
F_0^{(0)}, F_1^{(0)}, {\ldots} , F_k^{(0)}
\bigr\rbrack}
{W_{k+n} \bigl\lbrack F_0^{(1)}, F_1^{(1)}, {\ldots} , F_{n-1}^{(1)},
F_0^{(0)}, F_1^{(0)}, {\ldots} , F_{k-1}^{(0)}
\bigr\rbrack}\right)\right)_x-\frac{2x}{3}, \lab{y2nk}\\
\mu^2&=\left(-\frac{1}{3}+n\right)^2 ,\qquad \nu=2k-n-1  \nonumber\\
y^{(1,-n, -k)}_{-}&=\left(\ln \left(
\frac{W_{k+n+1} \bigl\lbrack 
{\widehat F}_0^{(1)},{\widehat F}_1^{(1)}, {\ldots} , {\widehat F}_{n-1}^{(1)},
{\widehat F}_0^{(0)}, {\widehat F}_1^{(0)}, {\ldots} , {\widehat F}_k^{(0)}
\bigr\rbrack}
{W_{k+n} \bigl\lbrack {\widehat F}_0^{(1)}, {\widehat F}_1^{(1)}, 
{\ldots} , {\widehat F}_{n-1}^{(1)},
{\widehat F}_0^{(0)}, {\widehat F}_1^{(0)}, {\ldots} , {\widehat F}_{k-1}^{(0)}
\bigr\rbrack}\right)\right)_x-\frac{2x}{3}, \lab{ym1nk}\\
\mu^2&=\left(-\frac{1}{3}+n\right)^2 ,\qquad \nu=-2k+n+1  \nonumber\\
y^{(2,-n, -k)}_{-}&=\left(\ln \left(
\frac{W_{k+n+1} \bigl\lbrack 
{\widehat F}_0^{(1)}, {\widehat F}_1^{(1)}, {\ldots} , {\widehat F}_{n-1}^{(1)},
{\widehat F}_0^{(2)}, {\widehat F}_1^{(2)}, {\ldots} , {\widehat F}_k^{(2)}
\bigr\rbrack}
{W_{k+n} \bigl\lbrack {\widehat F}_0^{(1)}, {\widehat F}_1^{(1)}, {\ldots} , 
{\widehat F}_{n-1}^{(1)},
{\widehat F}_0^{(2)}, {\widehat F}_1^{(2)}, {\ldots} , 
{\widehat F}_{k-1}^{(2)}
\bigr\rbrack}\right)\right)_x-\frac{2x}{3}, \lab{ym2nk}\\
\mu^2&=\left(\frac{1}{3}+n\right)^2 ,\qquad \nu=-2k+n-1  \nonumber
\end{align}
The corresponding symmetric form of the Painlev\'e equation
for the above solutions is e.g. 
\[
y^{(1,n, k)}_{+}+y^{(2,k+1, n+1)}_{+} + y^{(1,-(n-k), -(n+1))}_{-}
=-2 x - \left(\ln y^{(1,-(n-k), -(n+1))}_{-}\right)_x, \quad n \ge k
\]
and
\[
y^{(1,n, k)}_{+}+y^{(2,k+1, n+1)}_{+} + y^{(2,-(k-n), -k)}_{-}
=-2 x - \left(\ln y^{(2,-(k-n), -k)}_{-}\right)_x, \quad k \ge n \, .
\]
In view  of equation \rf{gnewdef} the above construction 
realized the B\"acklund transformations
as compositions of the DB and the
permutation transformations. 
Indeed, assigning $(v_i,v_j,v_k)= ((n+k+1)/3,(k-2n)/3,(n-2k-1)/3)$ to
the $y^{(1)}_{(k,n)}$ solution one verifies that the class of solutions 
shown in this subsection closes under the action of
the B\"acklund transformations $g_i,g_j, g_k$.
For example, the transformation $g_j$ maps 
$y^{(1)}_{(k,n)}$ to $y^{(2)}_{(k^{\prime},n^{\prime})}$
with $n^{\prime}=n-k, k^{\prime}=n$ for $n>k$ and
${\widehat y}^{(2)}_{(-k^{\prime},-n^{\prime})}$
with $n^{\prime}=k-n, k^{\prime}=k+1$ for $n<k$.

\vskip .4cm \noindent
{\bf Acknowledgements} \\
JFG and AHZ thank CNPq and FAPESP  for 
financial support.
Work of HA is partially supported by grant NSF PHY-0651694.
HA thanks Nick Spizzirri for discussions and correcting errors
in the previous versions of the manuscript.

\end{document}